\documentclass[%
 reprint,
 amsmath,amssymb,
 aps
]{revtex4-2}

\usepackage{graphicx}% Include figure files
\usepackage{gensymb}
\usepackage{dcolumn}% Align table columns on decimal point
\usepackage{bm}% bold math
\usepackage{amsmath}
\usepackage{amssymb}
\usepackage{cellspace}

\newcommand{\apjl}{ApJ}
\newcommand{\apjs}{ApJS}
\newcommand{\mnras}{MNRAS}
\newcommand{\aap}{A\&A}
\newcommand{\araa}{ARA\&A}
\newcommand{\pasp}{PASP}

\begin{document}

\preprint{APS/123-QED}

\title{X-ray Reflection from the Plunging Region of Black Hole Accretion Disks}% Force line breaks with \\

\author{Jameson Dong}
\thanks{Cahill Center for Astronomy and Astrophysics, California Institute of Technology, Pasadena, CA 91125, USA }
 \email{jdong2@caltech.edu}
%Lines break automatically or can be forced with \\
\author{Guglielmo Mastroserio}
 \thanks{Dipartimento di Fisica, Universit\`{a} degli Studi di Milano, Via Celoria 16, I-20133 Milano, Italy}
% \email{Second.Author@institution.edu}
\author{Javier A. Garc\'{i}a}
 \thanks{X-ray Astrophysics Laboratory, NASA Goddard Space Flight Center, Greenbelt, MD 20771, USA}
 \thanks{Cahill Center for Astronomy and Astrophysics, California Institute of Technology, Pasadena, CA 91125, USA }
 %\author{Andrew Mummery}
 %\affiliation{Oxford Theoretical Physics, Beecroft Building, Clarendon Laboratory, Parks Road, Oxford, OX1 3PU, UK }
 \author{Adam Ingram}
 \thanks{School of Mathematics, Statistics and Physics, Newcastle University, Herschel Building, Newcastle upon Tyne, NE1 7RU, UK }
\author{Edward Nathan}
 \thanks{Cahill Center for Astronomy and Astrophysics, California Institute of Technology, Pasadena, CA 91125, USA }
\author{Riley Connors}
 \thanks{Villanova University, Department of Physics, Villanova, PA 19085, USA }
\date{\today}% It is always \today, today,
             %  but any date may be explicitly specified
\begin{abstract}
\noindent Accretion around black holes is very often characterized by distinctive X-ray reflection features (mostly, iron inner-shell transitions), which arise due to the primary radiation being reprocessed by a dense and relatively colder medium, such as an accretion disk. Most reflection modeling assume that emission stops at the inner-most stable circular orbit (ISCO), and that for smaller radii---in the plunging region---the density drops and the accretion flow is far too ionized for efficient line production. We investigate the spectral features of the reflection in the plunging regions of optically-thick and geometrically-thin accretion disks around black holes. We show that  for cases in which the density profile is considered constant (as expected in highly magnetized flows), or in cases in which the disk density is high enough such that the ionization still allows line formation within the ISCO, there is a significant modification of the observed reflected spectrum. Consistent with previous studies, we found that the impact of the radiation reprocessed in the plunging region is stronger the lower the black hole spin, when the plunging region subtends a larger area. Likewise, as for the case of standard reflection modeling, the relativistic broadening of the iron line is more pronounced at low inclination, whereas the blueshift and relativistic beaming effect is dominant at high inclination. 
% We can also verify the velocity and density profile of GRMHD simulations by their reflection spectra, and probe the value of ISCO stress as well. 
We also tested the effects of various prescriptions of the stress at the ISCO radius on the reflection spectrum, and found that several of these cases appear to show line profiles distinct enough to be distinguishable with reasonably good quality observational data.   
\end{abstract}
\maketitle

\section{Introduction}
Black holes, characterized by their intense gravitational fields, are one of the most interesting entities in the Universe. They can be broadly categorized into stellar mass black holes, formed from the gravitational collapse of massive stars, and supermassive black holes, which reside at the centers of galaxies~\citep{Fryer_2001,Greene_2020}. Black holes of all masses are often seen accreting matter via an accretion disk~\citep{NT1973}. These accreting black holes, whether stellar mass or supermassive, are among the Universe's most luminous phenomena, showcasing a range of dynamic behaviors and interactions with their environment~\citep{NT1973}.
%Accreting black holes are among the Universe's most luminous and enigmatic phenomena. Central to these radiant structures is a supermassive black hole, surrounded by a dynamic accretion disk of matter~\citep{NT1973}.

Material within the disk is heated from within by turbulence~\citep{Steve1998}, and heated externally due to radiation absorption. The high temperature of the disk causes it to emit thermally in the UV and X-ray bands~\citep{SHIELDS1978}. Some of the thermal photons from the disk are thought to be Compton up-scattered by a region close to the black hole that is filled with hot plasma referred to as corona~\citep{TP1975}. A portion of these high-energy, Comptonised X-ray photons end up incident back upon the disk, where they are reprocessed with the disk atmosphere before re-emerging as the `reflected' emission~\citep{SUNYAEV1979}.
The geometry of the corona is still debated, with several competing interpretations (e.g. the truncated disk model \cite{Esin1997, Poutanen1997}, a sandwich/patchy corona \cite{Haardt1991, Stern1995}, the base-of-the-jet/lamp-post \cite{Markoff2005}).  
%This corona acts as the primary X-ray source of reflection, emitting high-energy X-rays that illuminate the accretion disk.   

%Reflection in this context refers to the process where radiation from the corona is reprocessed by the upper layer of the accretion disk, producing a characteristic energy spectrum dictated by the atomic physics, the dynamics of the disk, and the relativistic effects~\citep{Fabian2000}.
The reprocessing of the high-energy photons in the upper layers of the accretion disk produces a resultant reflection spectrum with characteristic features, such as the `Compton-hump', along with many atomic features. These features are produced in the local rest-frame of the disk, so the observed reflection spectrum has a signature smearing due to relativistic effects from the dynamics of the disk~\citep{Fabian2000}.

%Since the radiative properties from accretion disks are produced by various aforementioned processes, these radiation activity will be instrumental to our understanding to the accretion process and properties of the corona. 
In the X-ray band, the most prominent of component of the reflection spectrum is the iron $\rm Fe~K\alpha$ line at 6.4 keV. The broadening and distortion of the characteristic $\rm Fe~K\alpha$ line will give important insights to the disk structure and the strong gravitational effects around the central black hole~\citep{Matt1992,Merloni:2003ic}.
%In particular, modelling of the broadening and the distortion of the narrow emission lines, such as the iron $\rm Fe~K\alpha$ line emitted at 6.4 keV, is often used to determine the disk structure and the properties of the central black hole such as its spin (references).

The minimum radius where stable circular orbits exist is referred to as the innermost stable circular orbit (ISCO). The region inside of the ISCO and outside of the horizon is known as the plunging region. In this domain, due to the lack of rotational support, matter is on its terminal descent into the black hole, with the governing physics heavily influenced by gravitational forces. Most previous reflection spectroscopy studies have considered only reflected emission from outside of the ISCO, ignoring the plunging region. There are a number of reasons for this. First, classic disk models~\citet{NT1973,Shakura1973} assume a zero stress, or torque, boundary at the ISCO, meaning that matter in the plunging region falls into the black hole near the speed of light. This leads to the density in the plunging region being very low, which in turn means that the matter within it will be fully ionized by the radiation from the corona, such that it emits no atomic lines~\citep{Reynolds2013}. Second, because of gravitational redshift, emission from the plunging region is only significant for a very compact corona~\citep{Fabian2014}. However, studies have shown that if the density inside the ISCO is not as low as naively inferred from classic disk models, the plunging region can be visible, and can have significant impact on the estimation of black hole spin~\citep{REYNOLDS2003,relxil,Riaz2022}. Preliminary evaluations also showed that at low spins, ignoring the plunging region could introduce systematic error~\citep{Reynolds2008}.
There have been attempts to study the plunging region using different approaches.
\citet{Zhu2012} applied post-processing radiative transfer to simulation results in~\citep{Penna_2010}, discovering a high-energy power-law tail emitted from the plunging region. \citet{Wilkins2020} computed the time delays between the continuum emission and the reflected emission (reverberation lags) when the photons are also reflected from the plunging region. 

Recent advancements in accretion disk modelling have made it reasonable to re-consider the time-averaged reflection spectrum from the plunging region. By assuming a purely viscous stress through the plunging region, \citet{Potter2021} extended the classic Novikov-Thorne (NT) disk profiles~\citep{NT1973} to the event horizon. \citet{Mummery2023} analytically extended the same classical disk models, allowing for the effects of radiation pressure and arbitrary values of the ISCO stress. For particular values of the ISCO stress the \citet{Mummery2023} solutions reproduce the numerical results of \citet{Potter2021}. 
They showed that there is non-negligible density at the edge of the plunging region, thus re-processed emission from there could contribute features to the reflected spectrum. 
In addition, when the disk is in the presence of a strong poloidal magnetic field matter falls in with a velocity much less than the free-fall, this is known as a Magnetically Arrested Disk (MAD)~\citep{Narayan}. Simulations of the MAD case showed that inside ISCO, we have a relatively flat density profile~\citep{Scepi2024}, which will further enhance the reflection from the plunging region. More importantly, recent observations are potentially indicative of the existence of a MAD~\citep{Not_Crank}. Another observation on X-ray binary MAXI J1820+070 also indicated that the edge of the plunge region has an important contribution to the overall spectrum~\citep{Fabian2020}. While the different models offer valuable insights, they often differ in their predictions and interpretations. By comparing the reflection spectra produced under different conditions and models, we have the potential not only to distinguish between these competing theories but also to validate their accuracy and applicability. 
% where the emissivity is the highest when the corona is near. 
% The ground work on the X-ray side has also been laid, particularly by the Reltrans reflection model~\citep{Ingram2019}, which opened new avenues for a detailed exploration of X-ray reflection from the plunging region. 
% and briefly touched upon the time-averaged spectra. 

In this work, we will take a closer look at the spectra, utilizing the full radiation transfer model created by Garc\'ia et al.~\citep{Xillver}. By accounting for the reprocessed radiation in the plunging region, we can achieve a more accurate depiction of the reflection spectra. This recognition might shed new lights on the estimation of black hole spin, a challenge encountered in previous studies~\citep{disagree}. Such discrepancies highlight the importance of refining our methodologies and understanding the underlying physics to reconcile these divergent findings. 

The value of the ISCO stress parameter is a rich subject of ongoing research and discussion. The simulation of \citet{Hawley_2001} found that, contrary to the assumptions of NT, the stress in the disk material does not vanish at the ISCO; this has been discussed abundantly 
%literature has been produced over this topic
~\citep{Menou_2003,Noble_2010,Abramowicz2010,Andy2019}. As for now, most disk models leave the ISCO stress as a free parameter. 

In this paper, we will present our work in the following order. In Section II we introduce the setup of the black hole, its accretion disk, and the corona, and how they interact with the radiation field. In Section III we will compare the reflection spectrum from a constant density disk with and without the plunging region, and compare the reflection spectra from generalized NT disks with different velocity and ISCO stress. 
\label{Sec::intro}
\section{Structure of the Radiation-Disk Interaction}
\begin{figure*}
\includegraphics[width=\textwidth]{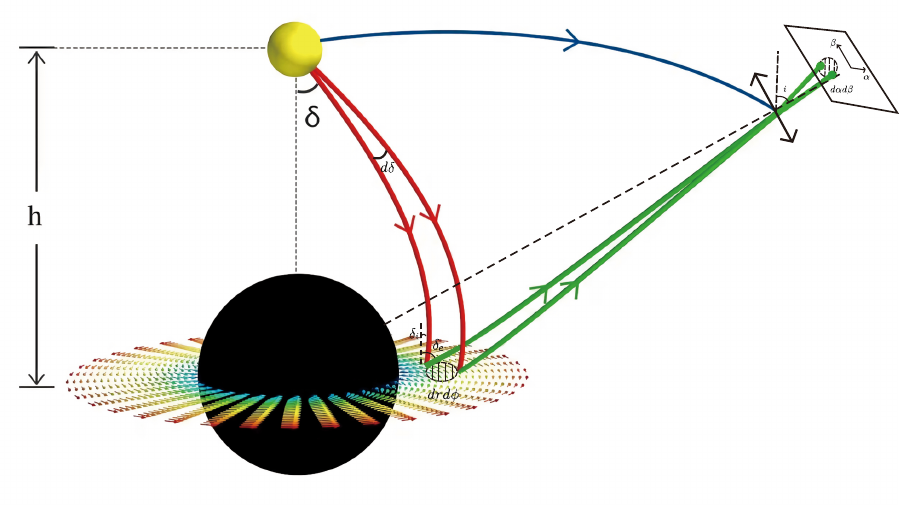}
\caption{Schematic of the rotating black hole, its accretion disk on in the equatorial plane and the on-axis lamp-post X-ray source. The observer is at the inclination angle of $i$. The red lines represent a beam of light rays emerged from the corona at a shooting angle of $\delta$, while spanning a incremental solid angle $d\Omega$. The beam shines on the disk patch $dA$, and the disk patch is cast to the small area of $d\alpha d\beta$ on the image plane of the observer. The blue line is the light ray goes directly from the corona to the observer.}
\label{scheme}
\end{figure*}
We work in the Kerr metric in the Boyer-Linquist coordinate, we adopt the natural unit where $GM=c=1$. The line segment in our spacetime can be described as
\begin{align}
g_{\mu\nu}dx^\mu dx^\nu = -\left(1 - \frac{2r}{\Sigma}\right)dt^2 - \frac{4a r\sin^2\theta}{\Sigma}dtd\phi\notag\\ + \frac{\Sigma}{\Delta}dr^2+ \Sigma d\theta^2  + \sin^2\theta\left(r^2 + a^2+ \frac{2a^2 r\sin^2\theta}{\Sigma}\right)d\phi^2,
\end{align}
where, $t$ denotes coordinate time as measured by a distant observer, $r$ is the radial distance from the black hole's singularity, adapted for curved spacetime. The angle $\theta$ measures deviation from the rotational axis, and $\phi$ represents azimuthal rotation around the black hole. Additionally, in the Kerr metric, we define $\Delta\equiv r^2-2r+a^2$ and $\Sigma \equiv r^2+a^2\cos^2\theta$.
\subsection{General  Description of the System}
In this paper, we evaluate the X-ray spectrum from an accreting black hole and its corona. In the scenario of our concern, the corona is modeled as a small spherical region on the rotation axis of the black hole which emits isotropically in its rest frame~\citep{Lampost, Matt1992}, and the accretion disk is optically thick and geometrically thin. \citet{Ingram2019} developed the \texttt{RELTRANS} model to calculate the disk reflection in this scenario up to ISCO. Here, we make use of their code and incorporate the necessary modifications to extend the calculation inside the ISCO radius. In Fig~\ref{scheme}, we depict the way in which the corona, the black hole, and the observer interact.
\subsection{ISCO Radius}
\citet{Bardeen1972} noted that not all circular orbits around the black hole are stable. When the effective potential $V(r)$ of a circular orbit (defined by Eqn.~2.10 in \citep{Bardeen1972}) has $V''(r)\leq 0$, a disturbance will cause orbiting matter to start inspiraling into the black hole. We can solve $V''(r)=0$ for the ISCO radius $r_I$
\begin{align}
    Z_1\equiv 1+(1-a^2)^{\frac{1}{3}}&\left( (1+a)^{\frac{1}{3}}+(1-a)^{\frac{1}{3}}\right)     \label{eq:isco1}\\
    Z_2\equiv(&3a^2+Z_1^2)^\frac{1}{2}\\
    r_I=3+Z_2\mp\big((3&-Z_1)(3+Z_1+2Z_2)\big)^\frac{1}{2}.
\end{align}
where the minus sign corresponds to prograde orbits and the plus sign corresponds to retrograde orbits.
\subsection{Velocity Profile and Lorentz Factors of the Accretion Disk }
\label{sec:velocity_profile}
For a geometrically thin and optically thick accretion disk, the orbit of a fluid element is considered to be Keplerian outside the ISCO, and its contra-variant 4-velocity will follow
\begin{equation}
   \boldsymbol{u}=u^{t}(1,0,0,\Omega_{\phi}),
   \label{eq:out}
\end{equation}
where 
\begin{align}
&\Omega_{\phi}\equiv\frac{1}{r^{3/2}+ a}, \\
u^{t}\equiv&\frac{r^{3/2}+ a}{\sqrt{r^2(r-3)+ 2a r^{3/2}}}.
%, which normalizes  $g_{\mu\nu}u^{\mu}u^{\nu}$ to  $-1$. 
\end{align}
When $r<r_I$, i.e. in the plunging region, we assume the fluid elements follow geodesics. The contra-variant 4-velocity of the fluid elements is
\begin{equation}
        u^t = \frac{1}{\Delta} \left( (r^2 + a^2 + \frac{2a^2}{r}) K - \frac{2aL}{r} \right)\label{eq:in1},
\end{equation}
\begin{equation}
    u^r = -\sqrt{\frac{2}{3r_{\text{\tiny I}}}}\left(\frac{r_{\text{\tiny I}}}{r}-1\right)^{\frac{3}{2}},
        \label{eq:in2}
\end{equation}
\begin{equation}
    u^{\theta}=0,
\end{equation}
\begin{equation}
        u^{\phi} = \frac{1}{\Delta} \left( \frac{2 a K}{r} + \left(1 - \frac{2}{r} \right) L\right),
        \label{eq:in3}
\end{equation}
where $k$ and $l$ are respectively the energy and the angular momentum of the disk patch at the ISCO, and they are defined as
\begin{align}
   &K=\left(1 - {\frac{2}{3r_I}}\right)^{1/2}, \\%&\frac{r_{\text{\tiny I}}^{\frac{3}{2}}-2r_{\text{\tiny I}}^{\frac{1}{2}}+a}{\sqrt{r_{\text{\tiny I}}^3-3r_{\text{\tiny I}}^2+2ar_{\text{\tiny I}}^\frac{3}{2}}} \\
    L&= 2\sqrt{3}  \left( 1 - {\frac{2a}{3 \sqrt{r_I}}}\right).%&\frac{r_{\text{\tiny I}}^2-2 a r_{\text{\tiny I}}^{\frac{1}{2}} +a^2}{\sqrt{r_{\text{\tiny I}}^3-3r_{\text{\tiny I}}^2+2ar_{\text{\tiny I}}^{\frac{3}{2}}}}
\end{align}
It is worth mentioning that Eqns. 5-7 converge to Eq.~\ref{eq:isco1} smoothly when $r\rightarrow r_{I}$, and the form of (9, 12 and 13) is recently simplified in~\citep{AndyPRL}.
In order to calculate the emissivity of a moving fluid element at a specific radius, we need to know its relativistic surface area contraction relative to a co-rotating observer at the same spot. 

The frame of such an observer is defined as a normalized local non-rotating frame (LNRF), where the new co-variant bases are:
\begin{align}
\boldsymbol{\hat{e}}^0&=\sqrt{-g_{tt}+\omega^2 g_{\phi\phi}} \boldsymbol{dt}\label{eq:LNRFbase1}\\
\boldsymbol{\hat{e}}^1&=\sqrt{g_{rr}} \boldsymbol{dr}\\
\boldsymbol{\hat{e}}^2&=\sqrt{g_{\theta\theta}} \boldsymbol{d\theta}\\
\boldsymbol{\hat{e}}^3&=-\omega\sqrt{g_{\phi\phi}} \boldsymbol{dt}+\sqrt{g_{\phi\phi}} \boldsymbol{d\phi} \label{eq:LNRFbase4}
\end{align}
and $\omega\equiv\frac{-g_{t\phi}}{g_{\phi\phi}}$. In the LNRF defined by base vectors \ref{eq:LNRFbase1}$\sim$\ref{eq:LNRFbase4}, the metric is $\boldsymbol{\eta}={\rm diag}\{-1,1,1,1\}$ \citep{Bardeen1972}.
We denote the 4-velocity of a disk patch in the LNRF as 
\begin{equation}
v^{\mu}\equiv(\boldsymbol{\hat{e}^{\mu}})_{\nu}u^{\nu}.
\label{Eq::v_in_LNRF}
\end{equation}
Since our frame is locally Minkowskian, we can calculate the Lorentz factor
\begin{equation}
\gamma=\frac{1}{\sqrt{1-(\frac{v^r}{v^t})^2-(\frac{v^\phi}{v^t})^2}}=v^t,
\end{equation}
the last equality is because we are in a locally flat coordinate.
Then, to a stationary observer at the lamp-post, differential element of the surface area of the disk annulus between $r$ and $r+dr$ becomes:
\begin{equation}
    dA_{\text \tiny Disk}=\gamma\ dA_{\text \tiny Kerr}=2\pi \gamma \sqrt{g_{\phi\phi}g_{rr}}dr 
\end{equation}
where $dA_{\text \tiny Kerr}$ is the area of the same annulus of the Kerr background.
\subsection{Light Rays and Energy-shift}
In this work, we denote the energy-shift of a photon from the source to the observer with a g-factor, $g_{so}=\frac{E_O}{E_S}$, where $E_S$ is the photon's initial energy in the rest frame of its source, and $E_O$ is the photon's final energy in the observer's rest frame. 

By covariantizing the special relativistic Doppler shift, we know that the energy for a photon with 4-momentum $k_\mu$ seen by an observer with 4-velocity $u^\mu$ is $E_O=-k_\mu u^\mu$.  For a photon travelling in the Kerr background, its co-variant 4-momentum is derived by Carter~\citep{Carter}
\begin{equation}
    \boldsymbol{k}=E(-1,\pm \sqrt{R}/\Delta,\pm \sqrt{q+a^2 \cos{\theta}^2},\lambda),
    \label{eq:4mom}
\end{equation}
where $E$ is the energy of the photon, since g-factor is a ratio of energies, we will omit $E$ from here on. $q$ is Carter's constant, a constant of motion in the Kerr spacetime, $\lambda$ is the photon's angular momentum around the z-axis, and the definition of $R$ is
\begin{equation}
    R=r^4 -(q+\lambda^2 -a^2)r^2+2r(q+(\lambda-a)^2)-a^2q,
    \label{eq::momentum}
\end{equation}
and the $\pm$ in Eqn.~\ref{eq::momentum} allows the photon to travel in either direction of the $r$ and $\theta$ coordinates.
\subsubsection{Lamp-post to Disk}
For our on-axis lamp-post at $r=h$, $\theta=0$. Since we are on the axis, and the emerging photons have $\lambda=0$, while the Carter constant can be written as
\begin{equation}
    q=-a^2+\frac{(a^2+h^2)^2 \sin^2\delta}{h^2-2h+a^2},
\end{equation}
where $\delta$, as shown in~Fig.\ref{scheme} stands for the angle between the initial direction of the light ray and the axis.

The contra-variant 4-velocity of the static source is 
\begin{equation}
   \boldsymbol{w}=\frac{1}{\sqrt{-g_{tt}}}(1,0,0,0)=\sqrt{\frac{h^2+a^2}{h^2-2h+a^2}}(1,0,0,0).
\end{equation}
The energy-shift of the photons observed by the disk is expressed as
\begin{equation}
    g_{sd}=\frac{k_\mu u^\mu}{k_\nu w^\nu}=\sqrt{\frac{h^2-2h+a^2}{h^2+a^2}}k_\mu u^\mu,
\end{equation}
where $u^\mu$ takes the form of Eqn.~\ref{eq:out} outside the ISCO and the form of Eqns.~\ref{eq:in1}--\ref{eq:in2} inside ISCO. We should pay extra attention to the sign of $k_{r}$, the radial momentum of the photon. It is possible that some light rays are already falling into the gravity well by the time they hit the disk inside the ISCO, so their  $k_{r}$ could be on the negative branch. 
\subsubsection{Disk to Observer}
\citet{Cunningham1973} observed that the 4-momentum of a photon, as it reaches a distant observer, is determined by the observer's line-of-sight inclination $i$ and the spatial intersection of this line with the photon's trajectory. This intersection in three-dimensional space can be described through the Cartesian coordinates of the photon's impact point on the observer's film, denoted by impact parameters $\alpha$ and $\beta$. This setup is depicted in Fig.~\ref{scheme}.  

When the observer is infinitely far away, Carter constant $q$ and angular momentum $\lambda$ of such photon is conveniently defined by $\alpha$ and $\beta$
\begin{align}
        q = \beta^2 &+ (\alpha^2 -a^2) \cos^2 {i}\\
        \lambda &= -\alpha \sin{i}
\end{align}
For a static observer infinitely far away from the system, the g-factor from the disk is:
\begin{equation}
    g_{do}=\frac{1}{k_\mu u^\mu}
    \label{name}
\end{equation}
%Inside the ISCO, the denominator has an additional term $k_r u^r$ from the radial components. For a given pair of impact parameters, we call YNOGK to trace which disk patch the photon emerged from. Similar to the lamp-post to disk case, we need to determine the sign of $k_r$ by ray-tracing. 
\subsubsection{Lamp-post to Observer}
For a model to be realistic, the direct spectrum from lamp-post is important. Since both the lamp-post and the observer are static, in this case, the only mechanism involved is gravity
\begin{equation}
        g_{so}=\sqrt{\frac{h^2-2h+a^2}{h^2+a^2}}.
\end{equation}
\subsection{Incident and Emission Angles}
To calculate the spectral flux density going into and coming out of the disk, we need to know the incident and emission angles of the light rays. Here we explain in detail how to calculate these angles of on given fluid element on the surface of the disk. The incident angle, defined as $\delta_i$ in Fig.~\ref{scheme}, is the angle between the normal vector of the disk patch and the light ray originated from the corona in the rest frame of the disk patch. The emission angle, defined as $\delta_e$, is the angle between the normal vector of the disk patch and the light ray destined to the observer in the aforementioned frame. 

We proceed in the LNRF, because its base tetrad Eq. \ref{eq:LNRFbase1}$\sim$\ref{eq:LNRFbase4} leads to a locally Minkowsikian metric $\boldsymbol{\eta}={\rm diag}\{-1,1,1,1\}$. For a disk patch at radius $r$ and moving through the spacetime with 4-velocity $\boldsymbol{v}$ given by Eq. \ref{Eq::v_in_LNRF}, we do a Lorentz transformation $\Lambda^\mu_\nu$ into its rest frame: 
\begin{equation}
	\boldsymbol{v'}=\boldsymbol{\Lambda  v}=(1,0,0,0).
\end{equation}

We denote all the quantities in the fluid element frame by a prime.  Since$~\boldsymbol{\Lambda}$ is Lorentzian, our new metric remains locally Minkowskian, $\boldsymbol{\eta '}=\boldsymbol{\eta}={\rm diag}\{-1,1,1,1\}$

In the LNRF, the equatorial plane where the accretion disk resides is a 3-hypersurface $\theta=\frac{\pi}{2}$ with normal vector $\boldsymbol{\hat{n}} = (0,0,1,0)$. 
 Since $u^\theta=v^2=0$, this Lorentz transformation does not involve $\theta$ and the space-like normal vector of the moving disk patch remains the same, $\boldsymbol{\hat{n}'} =\boldsymbol{\Lambda  n}= (0,0,1,0)$.
 
The last piece we need is the 4-momentum of the incoming and out-going photons in the rest frame of the disk patch. Similarly, we transform Eq. \ref{eq:4mom} first to LNRF then to the patch frame
\begin{equation}
    k'^{\eta}\equiv\Lambda^\eta_\mu(\boldsymbol{\hat{e}^{\mu}})_{\nu}k^{\nu}
\end{equation}
We now have all the relevant 4-vectors in the patch frame, to evaluate the spatial angles, we project the vectors (and the metric) onto the time slice when the photon is absorbed or emitted $t'=t_{0}$ and calculate their inner product for the cosine of the spatial angles. 

For the projection, to define a (trivial) projector $\boldsymbol{P}$ that projects 4-vectors onto the time slice
\begin{equation}
P^i_\mu \equiv\frac{\partial x'^i}{\partial x'^\mu},
\end{equation}
where $x^\mu$ are the 4-coordinates in the patch frame, and $i$ (and other Latin characters) are the spatial indices running from 1 to 3~\citep{poisson}. And we define the induced inverse metric on the time slice as $h^{ij} \equiv P^i_\mu P^j_\nu \eta'^{\mu\nu}=\delta^{ij}$. \\
Finally, we get to the incident angle with the spatial inner product on the time slice:
\begin{equation}
    \cos{\delta_i}=\frac{(\boldsymbol{P\hat{n}'})\cdot(\boldsymbol{Pk'})}{\sqrt{|\boldsymbol{P\hat{n}'}||\boldsymbol{Pk'}|}}=\frac{h_{ij}P^i_\mu n'^\mu P^j_\nu k'^\nu}{\sqrt{|\boldsymbol{P\hat{n}'}||\boldsymbol{Pk'}|}}
\end{equation}
\newcommand{\nprime}{{n'}}
\newcommand{\kprime}{{k'}}

In the numerator, since the only non-zero term in $\boldsymbol{\hat{n}'}$ is $\hat{n}'^2$, the summation reduces to 
\begin{equation}
   h_{22}P^2_2{\nprime}^2P^2_2{\kprime}^2={\kprime}^2=(\boldsymbol{\hat{e}_2})^\nu k_\nu=\frac{\sqrt{q}}{r}. 
\end{equation}
 In the denominator, since $\boldsymbol{Pk'}=({\kprime}_1,{\kprime}_2,{\kprime}_3)$, and in a Minkowski background, $|\boldsymbol{Pk'}|=\sqrt{({\kprime}_1)^2+({\kprime}_2)^2+({\kprime}_3)^2}$. The normalization of a photon 4-momentum is $-({\kprime}_0)^2+({\kprime}_1)^2+({\kprime}_2)^2+({\kprime}_3)^2=0$, so $|\boldsymbol{Pk'}|={\kprime}_0=u^\mu k_\mu$. We have recovered the expression for the cosine of the incident angle
\begin{equation}
	\cos{\delta_i}=\frac{\frac{\sqrt{q}}{r}}{u^\mu k_\mu},
\end{equation}
and the calculation for the emission angle is similar.
\subsection{Spectrum Components, Luminosity, and Emissivity}
In this subsection, we will follow \citet{Ingram2019} and briefly introduce the calculation for the components of the total spectrum. 

For an isotropic lamp-post with surface area $S$, We can define its specific flux as:
\begin{equation}
    F_{S}(E,t)=\frac{C(t)}{S}f(E|\Gamma,E_{\rm cut})
    \label{eq::source}
\end{equation}
where $C(t)$ describes time variability of the source, though this work only study time-averaged spectrum, we keep it for the sake of generality, and 
\begin{equation}
    f(E|\Gamma,E_{cut})\propto E^{1-\Gamma}e^{\frac{-E}{E_{\rm cut}}},
\end{equation}
is the assumed incident spectral shape with photon index $\Gamma$.

From Eqn.~\ref{eq::source} we also have the direct specific flux from the lamp-post seen by an observer at a distance $D$ and cosmological redshift $z$
\begin{equation}
    F_{D}(E,t)=\frac{C(t-\tau_{so})}{4\pi D}\left(\frac{g_{so}}{1+z}\right)^\Gamma f(E|\Gamma,g_{so}E_{cut}/(1+z)),
\end{equation}
where $\tau_{so}$ is the time for a photon to travel to the observer.

For the reflection spectrum, we define the specific intensity as 
\begin{equation}
dF_{R}(E,t)=g_{do}^3(r,\phi) I\left(\frac{E}{g_{do}(r,\phi)},t-\tau_{do}(r,\phi)\right)\frac{d\alpha d\beta}{D^2}
\end{equation}
The mapping between the impact parameters and coordinates on the disk, $\{r(\alpha,\beta),\phi(\alpha,\beta)\}$ will be provided by the ray-tracing library YNOGK~\citep{Yang2013}. $I$ is the specific intensity emitted by the illuminated disk:
\begin{equation}
I(E,t,\mu_{e})=\frac{1}{2}C(t-\tau_{sd})g_{sd}^\Gamma\frac{R(E,g_{sd}E_{cut},\delta_{e})}{4\pi}\frac{|d\cos{\delta}|}{dA_{\rm ring}}
\label{Eq::flux}
\end{equation}
where $dA_{\rm ring}$ is the area of the annulus between the impact radii of photons from the lamp-post with shooting angle $\delta$ and $\delta+d\delta$. And $R(E,g_{sd}E_{cut},\delta_{e})$ is the specific intensity calculated by our reflection model \texttt{XILLVER}, which we will briefly introduce in the next subsection.

Since our reflection model assumes equal in-going and out-going radiation power, we can define the disk radial emissivity profile by the specific flux density of the radiation from the lamp-post:
\begin{equation}
    \epsilon(r)=\frac{g^{\Gamma}_{sd}(r)}{2}\frac{|d\cos{\delta}|}{dA_{\rm ring}}.
\end{equation}
\subsection{Reflection Model}
In this subsection we briefly describe the plane-parallel reflection model \texttt{Xillver} that produces the reflection function $R(E1,E2,\mu_e)$ in Eq.~\ref{Eq::flux}. A more detailed discussion is can be found in~\citep{Garcia2013}. 
When radiation is illuminating a plane-parallel gas, the transfer of the radiation field is described by
\begin{equation}
    \mu^2 \frac{\partial^2 \xi(\mu, E, \tau)}{\partial \tau^2} = \xi(\mu, E, \tau) - S(E, \tau),
    \label{eq::rad}
\end{equation}
where $ S(E, \tau )$ is the source function in the atmosphere, and $\xi(\mu, E, \tau)$ is the specific intensity of the radiation field for
propagating with cosine value $\mu$ relative to the normal direction of the atmosphere, $\tau$ is the total optical depth, whose differential is defined by spatial depth $x$ into the atmosphere and total opacity $\chi(E)$,
\begin{equation}
    d\tau=\chi(E)dx.
\end{equation}
\texttt{Xillver} solves Eqn.~\ref{eq::rad} with the top and bottom boundary conditions,
\begin{equation}
    \mu \left[ \frac{\partial \xi(\tau, \mu, E)}{\partial \tau} \right]_{0} - \xi(0, \mu, E) = -\frac{F_{in}}{\cos{\delta_i}} \delta(\mu - \cos{\delta_i}),
    \label{eq::top}
\end{equation}
\begin{equation}
    \mu \left[ \frac{\partial \xi(\tau, \mu, E)}{\partial \tau} \right]_{\tau_{\max}} + \xi(\tau_{\max}, \mu, E) = B(T_{\text{disk}}),
    \label{eq::bottom}
\end{equation}
where $B(T)$ is the Planck function and $T_{disk}$ is the temperature of the disk, and $F_{in}$ is the flux from the corona seen in the frame of the fluid element, 
\begin{equation}
    F_{in}(E_d, t') = \frac{C(t' - \tau_{sd})}{4\pi} g_{sd}^\Gamma f(E_d| \Gamma, g_{sd} E_{cut}) \frac{d \Omega_d}{d A_d}.
\end{equation}
After \texttt{Xillver} solved Eqn.~\ref{eq::rad}, it will update $S(E,\tau)$ by calling \texttt{Xstar}~\citep{xstar}, and the process will be repeated until both $\xi(\mu, E, \tau)$ and $S(E, \tau)$ converge.

Then, $R(E,g_{sd}E_{cut},\delta_{e})=\xi(0, \cos{\delta_{e}}, E)$ at the end of all iterations.

\section{Results and Discussion}
In this section, we will deliver the following results. First we show that the brightest part of the accretion disk lies in the plunging region. Then, assuming a constant density disk, we compare the reflection spectra with and without the plunging region at different black hole spins and observer inclinations. Finally, utilizing disk models provided by~\cite{Mummery2023}, we examine the spectroscopic impacts of varying the velocity profile and ISCO stress. The list the parameters involved in our reflection model is shown in Table~\ref{tab:par}. Unless mentioned otherwise, we use their default values.
\\
\begin{table}[ht]
\centering
\begin{tabular}{|p{1.5cm}|p{1.5cm}|p{3.7cm}|p{1.1cm}|}
\hline
\small{Parameter} & \small{Units} &\small{Description} & \small{Default Value} \\
\hline
$M$ & $M_\odot$ &\small{Black hole mass}& $10^7$\\
\hline
$a$ & $ $ &\small{Black hole dimensionless spin}& 0\\
\hline
$h$ & $\frac{GM}{c^2}$ & \small{Source height} & 6 \\[1ex]
\hline
$r_{in}$ &$r_{horizon}$ & \small{Disk inner radius} & 1 \\
\hline
$r_{out}$ &$\frac{GM}{c^2}$ & \small{Disk outer radius} & 1000 \\
\hline
$z$ &  & \small{Cosmological redshift} & 0\\
\hline
$\Gamma$ &  & \small{Photon index} & 2\\
\hline
$\xi_{ion}$ & \small{$\rm erg~cm~s^{-2}$} & \footnotesize{Ionization parameter at ISCO} & $\rm 10^3$\\
\hline
$A_{Fe}$ & \small{Solar} &\footnotesize{Relative iron abundance}& \\
\hline
$E_{cut}$ & \small{keV} &\footnotesize{Observed high energy 
cut-off}& 60\\
%$kT_e$ & \small{keV} &\footnotesize{Observed}& 60\\
%&& \footnotesize{electron temperature}&\\[-2ex]
\hline
$\small{\dot{M}/\dot{M}_{Edd}}$ &  &\footnotesize{Mass accretion rate}& 0.1\\
\hline
$\alpha$& &\footnotesize{Viscosity parameter}&0.1\\
\hline
$\delta_J$& &\footnotesize{ISCO stress parameter}&0.1\\
\hline
\end{tabular}
\label{tab:default_par}
\caption{Default parameters used for the reflection calculation}\label{tab:par}
\end{table}

\subsection{Radial Emissivity Profile and Relative Flux Contribution}
\begin{figure}
\centering
\includegraphics[width=0.95\columnwidth]{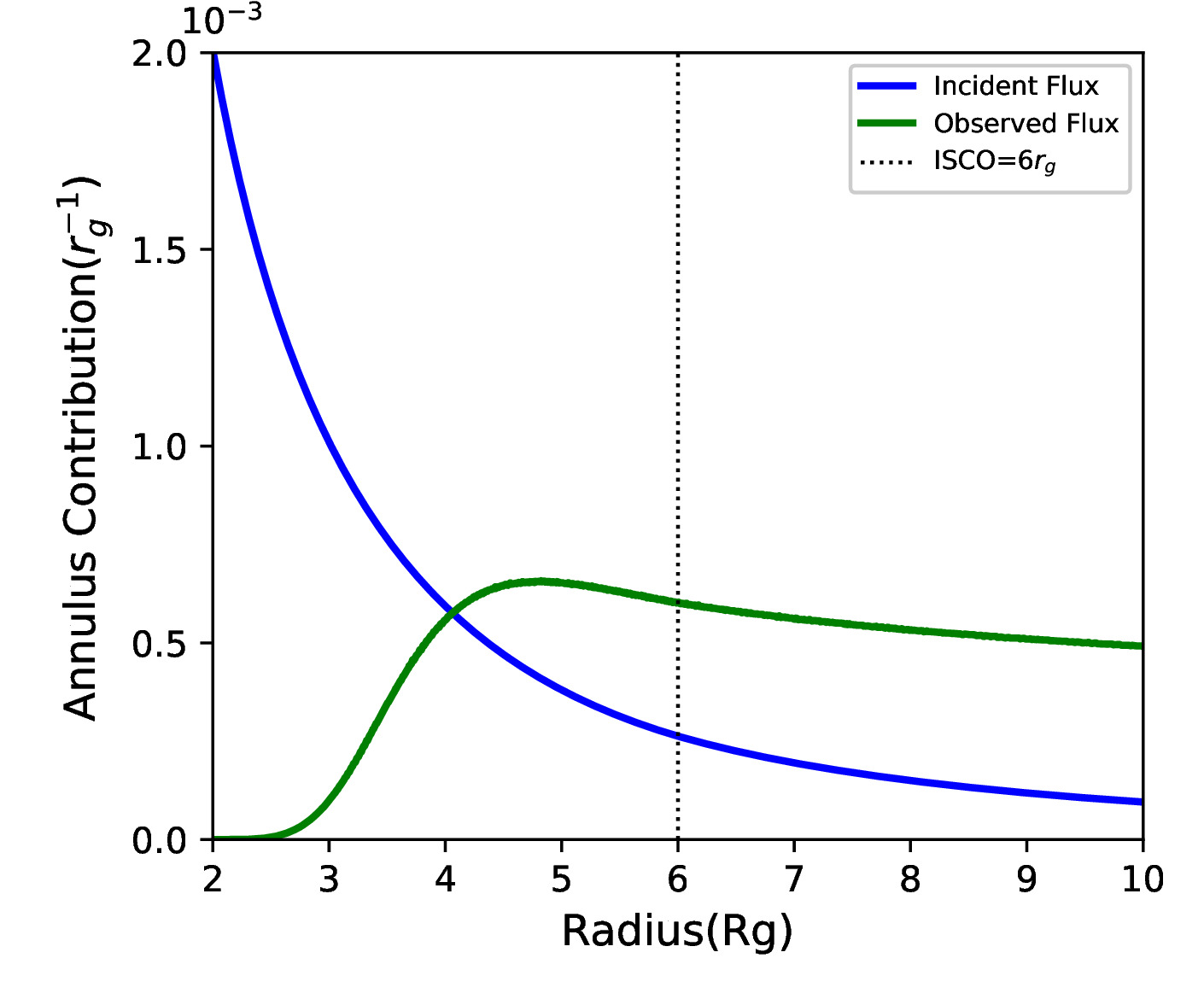}% Here is how to import EPS art
\caption{Fraction of the total incident (blue) and observed (green) flux from an unit width disk annulus at $r$. The black hole is Schwarzschild ($a=0$) and the inclination angle of the observer is $75\degree$.}
\label{fig:relative}
\end{figure}

We first show the contribution of the plunging region in terms of incident and observed flux. We consider a non-spinning black hole ($a=0$) and inclination $75\degree$, and the rest of the parameters are set to their default values (listed in Table~\ref{tab:par}). Figure~\ref{fig:relative} shows the incident and observed fluxes at an unit width annulus at $r$, expressed in terms of the fractional contribution to the total flux. 

The incident flux for a unit width annulus at $r$ can be defined by the annulus' solid angle to the corona
\begin{equation}
    \zeta(r)=\frac{g^{\Gamma}_{sd}(r)}{2}|\frac{d\cos{\delta}}{dr}|.
\end{equation}

To gauge the observed flux contribution across all radii, we define flux that comes from a unit width annulus with inner radius $r$ and reaches the observer
\begin{equation}
     \kappa(r)=\frac{1}{\Delta r}\int_{r'=r}^{r'=r+\Delta r}g_{do}(r',\phi)^3\epsilon(r')dA_{ring},
     %\Delta r\int_{r'=r_{in}}^{r'=r_{out}}g_{do}(r',\phi)^3\epsilon(r')dA_{ring},
\end{equation}
where $\Delta r$ is a small increment in radius. We normalize $\zeta(r)$ and $\kappa(r)$ by their integration over the whole disk and examine their behavior in Fig.~\ref{fig:relative}. We note that, although the disk in the vicinity of the event horizon receives the most flux from the corona, to the observer, the brightest part of the disk lies between $4r_g$ and $5r_g$. 

The explanation for this is that the majority of the photons emitted within 3 $R_{\rm g}$ cannot escape the gravitational pull of the black hole and never reach the observer. 

Of course, for higher values of the black hole spin the ISCO is smaller, and therefore the plunging region becomes increasingly smaller too. We then do not expect much contribution from the plunging region when the spin is maximal ($a=0.998$) or close to maximal.

\subsection{Constant Density Accretion Disks}
We choose to investigate accretion disks with constant density profiles. This choice is motivated by the fact that some MAD simulations exhibit near-constant density profiles~\citep{MAD}. With an abundance of matter in the plunging region, this model accentuates reflections from this area, providing insight into the potential upper limit of the contribution from the plunging region.

\begin{figure*}
\includegraphics[width=\textwidth]{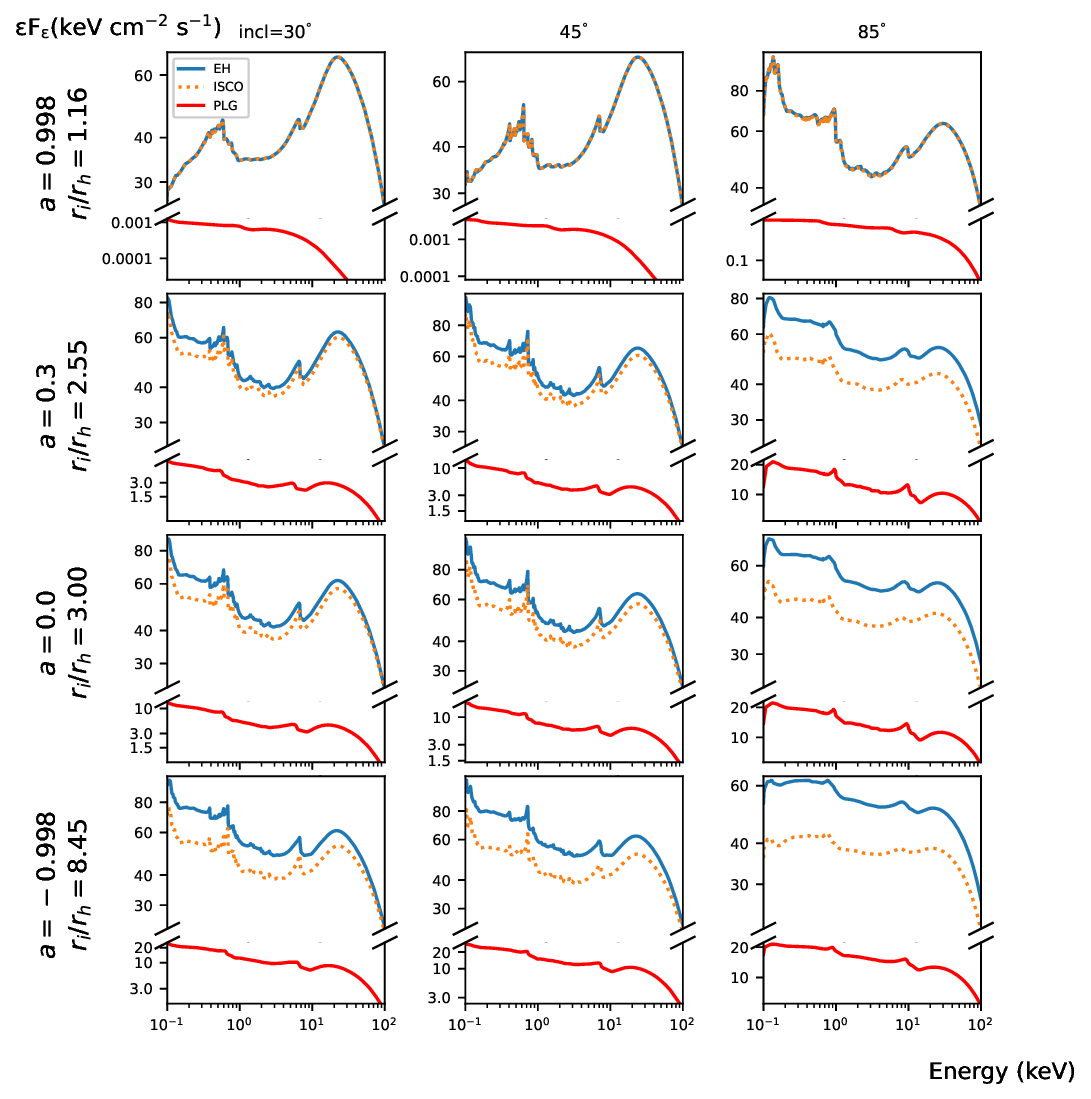}% Here is how to import EPS art
\caption{X-ray reflection spectra in the $0.1-100$\,keV energy range, for calculations assuming reflection occurring in the plunging region all the wau to the event horizon (blue solid line, labelled ``EH'') and for reflection stopping at the ISCO (orange dotted line, labelled ``ISCO''). The red solid line (labelled ``PLG'') represents the contribution to the reflected flux from the plunging region alone. Each column show a given inclination (increasing to the right), while each row show a different spin parameter (increasing to the bottom). The disk density is set to $\rm 10^{17} cm^{-3}$.}
\label{fig:AllSpec}
\end{figure*}

In Figure~\ref{fig:AllSpec} we compare the reflected X-ray flux calculated for a disk that stops at the ISCO (orange dashed lines) with the case of an accretion flow that produces reflection within the plunging region until the event horizon (blue solid lines). We also show the contribution of the reflection from the plunging region alone (red solid line). The spin value increases from the bottom row (retrograde case) to the top row (maximally spinning case), whereas the inclination angle increases from the left to the right column.

The contribution of the plunging region to the total integrated reflection spectrum mainly depends on the black hole spin (as it determines its total area), and the inclination (as it enhances the distortion of spectral features due to general relativistic effects). As shown in Figure~\ref{fig:AllSpec}, we found that in general the effect of the plunging region is larger at low spins and high inclinations, affecting the entirety of the reflected signal at all energies.

We note that the spin value positions the distance of the ISCO from the black hole, and thus sets the dimensions of the plunging region. Specifically, at maximum spin ($a=0.998$), when the ISCO is extremely close to the event horizon ($r_{\rm ISCO}/r_{\rm h} = 1.16$), we found that the contribution of the plunging region to the reflection signal is less than 1\% even at high inclinations. However, this contribution increases substantially as the spin parameter is lowered. Evidently, the most extreme case is found at maximum retrograde spin ($a=-0.998$; bottom row), where the emission from the plunging region contributes between $\sim25-33$\% of the total reflected spectrum at inclinations in the range $30-85\degree$.
We also noted that the inclusion of the plunging region changes the overall slope of the reflection spectrum, making it softer in all the cases.

 %We also note that at higher inclinations the reflected flux from the plunging region increases and the the reflection features become sharper. This happens because at high inclinations, the fluid velocity is more aligned with the observer's line-of-sight, thus the relativistic beaming effect becomes more pronounced. The photon number emitted from the parts of the plunging region which is moving away from or towards the observer is vastly reduced. Coupled with strong red-shift, plunging region flux is dominated by the blue-shifted part of the plunging region. We will discuss this phenomenon in detail in Section~\ref{subsec:ChopDisk}.

\begin{figure*}

\includegraphics[width=\textwidth]{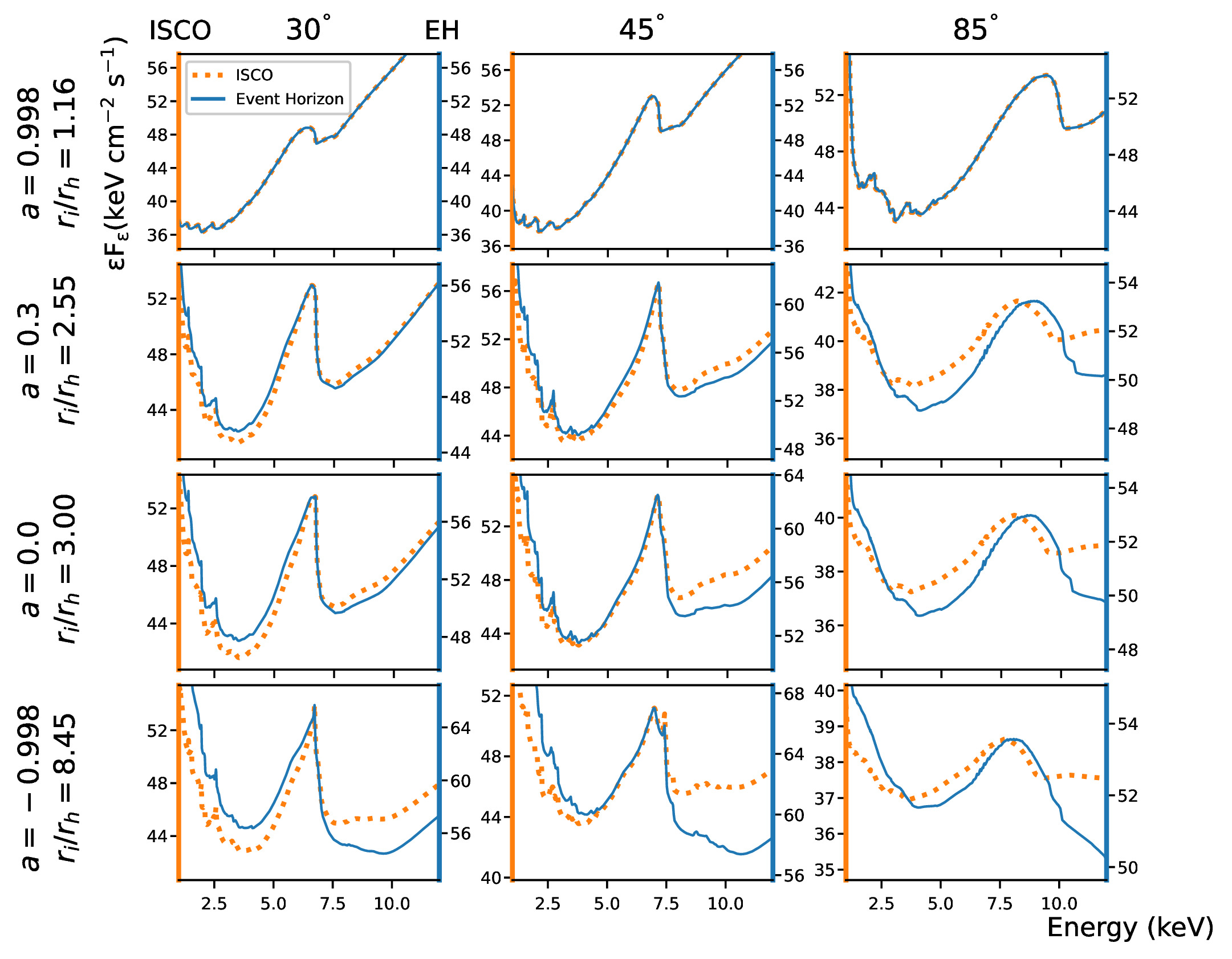}
\caption{A comparison of the $\rm Fe~K\alpha$ line spectral flux density, spanning a range of 1 -- 12 keV, is depicted for simulations from the event horizon (blue solid line) and the ISCO (orange dotted line). The parameters are the same as Fig~\ref{fig:AllSpec}. For the ease of comparison, we used dual y-axis to align the height of the spectral peaks. We can observe that, at low inclination, the GR broadening effect of the $\rm Fe~K\alpha$ line is the most pronounced, while at high inclination angles, the blueshift of the peak is more significant.}
\label{fig:K_Alpha}
\end{figure*}

Given that the $\rm Fe~K\alpha$ line is one of the most important features in reflection spectroscopy, particularly for the black hole spin determination, it is important to understand the impact of the contribution from the plunging region in the line profile. In Figure~\ref{fig:K_Alpha}, we zoom in on the $\rm Fe~K\alpha$ region alone and compare the flux density spectrum computed with and without considering the plunging region. The only difference (with respect to Fig.~\ref{fig:AllSpec}) is that the two lines in each panel refer to separate y-axes (color-coded left and right) to align the peaks of the lines and compare their shape. We found that at spins of $a=0.3$ or lower, the effects of the plunging region are important enough to produce appreciable modifications to the line profile, particularly at the highest inclinations, where significant blueshifts are observed. At low inclination, additional GR broadening effects can be observed on the red wing of the $\rm Fe~K\alpha$ line, whose slope is reduced. At medium inclination, the plunging region does not contribute to the red wing of the line, while it increases the slope of the blue wing.
At high inclination, the shape of the iron line looks steeper on both red and blue wings, and the overall profile peaks more towards higher energies.

In summary, the observed changes in the $\rm Fe~K\alpha$ line profile, and in the overall slope of the reflection spectrum, could have important implications for the interpretation of observational data. For example, the flux contributed by the plunging region produces a softening of the overall reflection spectrum continuum, which could affect the overall model parameters that describe the X-ray corona. Additional flux at soft energies could have other effects, such as a better description of the soft excess in AGN~\citep{Garcia_2019}. 
% or produce changes in the recovered iron abundance~\citep{garcia2018,Tomsick_2018}. 
Likewise, the changes observed in the $\rm Fe~K\alpha$ profile could impact the measurements of other parameters such as the iron abundance and the black hole spin. Regarding the numerous former works that reported a requirement for super-solar iron abundances to fit the reflection spectrum, here we have found that the plunging region contribution seems to increase the strength of the line, especially at high inclination. This effect may therefore subsume the requirement for higher iron abundances to fit the data. Similarly, the measurement of the black hole spin is likely to be affected by the plunging region contribution for the most extreme cases. 
% alone suggest that a system with low spin and high inclination with strong contribution from the plunging region could produce a line profile similar to those at higher spins (i.e., given the strong gravitational distortion suffered by the material closest to the black hole). This could have serious consequences in biasing the spin measured in some systems. 
A proper assessment of these effects in model fitting is outside the scope of this paper and will be addressed in future publications.

%While at high inclination, the blueshift $\rm Fe~K\alpha$ line profile is enhanced by the inclusion of the plunging region. We will discuss this in detail in Section~\ref{discuss::constant line profile}.  We note that the difference between the peak of the $\rm Fe~K\alpha$ line and the continuum next to the blue wing increases with the inclination angle of the observer. This happens because at higher inclination angles we observe stronger GR effects (as described in~\citep{Garcia_2014}), and it is even more pronounced when we include the plunging region.
%If we zoom into the range around the $\rm Fe~K\alpha$ line, in Fig~\ref{fig:K_Alpha}, we can make the additional observation that, in the presence of the plunging region, the difference between the peak of the $\rm Fe~K\alpha$ line and the continuum on its blue wing increase with the inclination angle of the observer. This difference has been described by Garcia et al.~\citep{Garcia_2014}, the higher inclination angles make the GR effects more observable, which compounds with the strong GR effects in the plunging region.


\subsection{Contribution from Different Sections of the Plunging Region}\label{subsec:ChopDisk}

\begin{figure*}
\includegraphics[width=\textwidth]{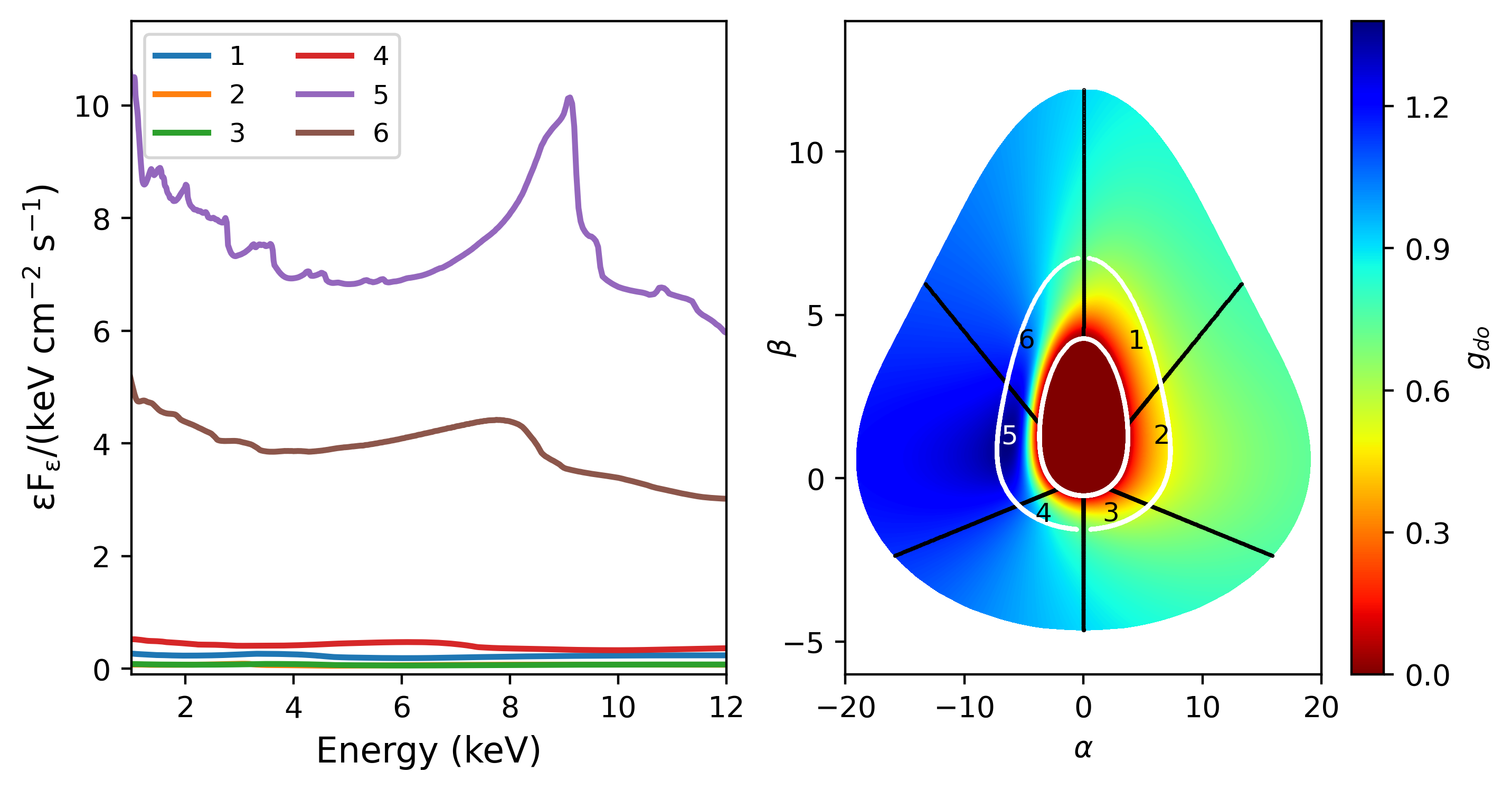}
\caption{A comparison of plunging region reflection flux density across different parts of the region. We are observing a Schwarzschild black hole at $\rm 75^\circ$ inclination. The right panel depicts the disk-to-observer redshift factor distribution on the observer's image. The outer white contour is the projection of ISCO on the observer's image, the inner white contour is the event horizon. The black lines divide the plunging region into 6 parts. For region 1, $\theta_o\in(0,\pi/6]$, region 2, $\theta_o\in(\pi/6,2\pi/3]$, region 3, $\theta_o\in(2\pi/3,\pi]$, region 4, $\theta_o\in(\pi,4\pi/3]$, region 5, $\theta_o\in(4\pi/3,11\pi/6]$, and region 6, $\theta_o\in(11\pi/6,2\pi]$. The right panel shows their respective flux density. }
\label{fig:chop}
\end{figure*}

To better understand the contribution of plunging region reflection, we isolate the reflection from different parts of the plunging region. We divide the plunging region by the azimuth on the observer's image $\theta_o$ defined in~\citep{Cunningham1973}. We aim to isolate distinct line features to evaluate their respective impact on the total spectrum. In Fig.~\ref{fig:chop} we show the image of the accretion disk on the observer camera in terms of impact parameters, the color scheme represents the energy shift applied to the photons emitted from the disk to the observer (right panel). 
Even though the image shows a portion of the disk outside the plunging region, the flux density spectra on the left-hand side of Fig.~\ref{fig:chop} are computed only considering the radiation emitted from the plunging region (numbered regions inside the white line). 
We note that, at high inclination ($75\degree$), the disk regions that are moving towards the observer (regions 5 and 6) contribute significantly more to the plunging region spectrum compared to all the other regions (1, 2, 3, and 4). This is due to the strong relativistic beaming effect at high inclination, which is accounted for by the GR reciprocity theorem in~\citep{reciprocity}, and reflected in our calculation by the fact that the reflection flux scales with $g_{do}^3$. In Fig.~\ref{fig:chop}, the azimuthal velocity of region 5 aligns with the direction to the observer, so the relativistic beaming is stronger than the other regions, and amplifies the observed flux. In region 6, it is the combination of radial and azimuthal velocity that acts similarly. In the other parts of the plunging region, the opposite is true: when the emitting fluid is moving away from the observer, the photon count at the observer decreases, and the photons are subject to severe redshift. 

\begin{figure*}
\includegraphics[width=\textwidth]{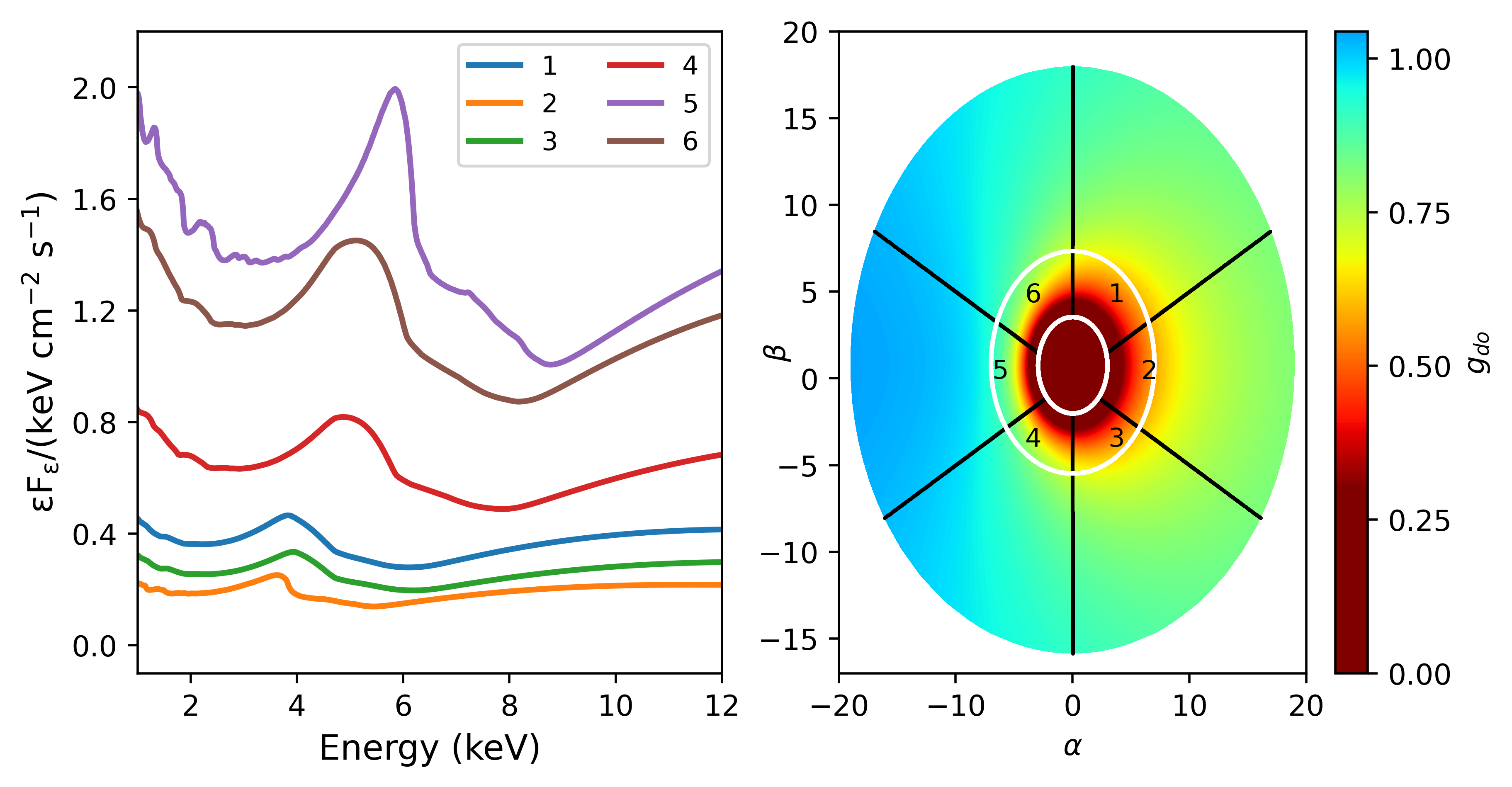}
\caption{The same comparison of plunging region reflection flux density across different parts. In this case, we observe a Schwarzschild black hole at $\rm 30^\circ$ inclination. In this figure, each region occupies an angle of $2\pi/3$ in $\theta_o$.}
\label{fig:chop2}
\end{figure*}

On the other hand, when the inclination is low, as Fig.~\ref{fig:chop2} depicts, the distribution of flux density across the plunging region is more balanced. 
However, the overall flux density from the plunging region is much lower for low inclinations, since no disk region is impacted by the extreme relativistic beaming towards the observer, as in the high inclination case. 
% moves towards the observer fast enough to counter the gravitational redshift, which in turn capped the GR broadening effects on the total spectrum. 
In the low inclination case, we can appreciate the different spectral shapes around the iron line region. 
We note that the $\rm Fe~K\alpha$ is always red-shifted with respect to its rest-frame emission energy (6.4~keV). Even in region 5, which includes the portion of the plunging region that rotates towards the observer, the gravitational red-shift dominates the blue-shift due to the azimuthal velocity and relativistic beaming. 
In the most extreme case of region 2 where the velocity of the disk points in the opposite direction to the observer, the $\rm Fe~K\alpha$ peaks below 4~keV. 
Even though in this case the plunging region contributes around 10\% of the flux (see third row, first column panel in Fig.~\ref{fig:AllSpec}), the red wing of the $\rm Fe~K\alpha$ 
 line is broader (same panel in Fig.~\ref{fig:K_Alpha}).

\subsection{Generalized Novikov-Thorne Disks}

As we mentioned in Section~\ref{Sec::intro}, \citet{Mummery2023} analytically extended the Novikov-Thorne disk model to the event horizon. While Novikov and Thorne assumes the stress at ISCO vanishes, \citep{Mummery2023} leaves ISCO stress as a free parameter. In this subsection, we move away from using a constant density profile within the plunging region. The bottom left panel of Fig.~\ref{fig:velocity} is a good example of the density profile of the generalized NT disk. And the velocity profile in~\citep{Mummery2023} has a small deviation from the geodesic. We now implement this extended NT disk into our reflection models, and discuss the spectroscopic implications.

\subsubsection{Impact of the non-Geodesic Velocity Profile in the Plunging Region}

We have so far assumed that material within the plunging region free-falls along the geodesics (see section~\ref{sec:velocity_profile}). Eqn.~\ref{eq:in2} indicates that the radial velocity of the accreting material is zero at the ISCO  ($u^r_I\equiv u^r(r_I)=0$).  However, as noted by \citet{Mummery2023}, this is unphysical as material has to have some velocity in order to cross the ISCO, and therefore real accretion flows should have a non-zero (if small) radial velocity at the ISCO. 

To consider the effect of the $u^r_I=0$ assumption, we compare the spectra generated from material with the velocity profile described by Eqns.~\ref{eq:in1}--\ref{eq:in3}, 
%which assume zero radial velocity at the ISCO, 
and the spectra generated considering the new assumption on the radial velocity profile from~\citep{Mummery2023}.

\begin{figure*}
\includegraphics[width=\textwidth]{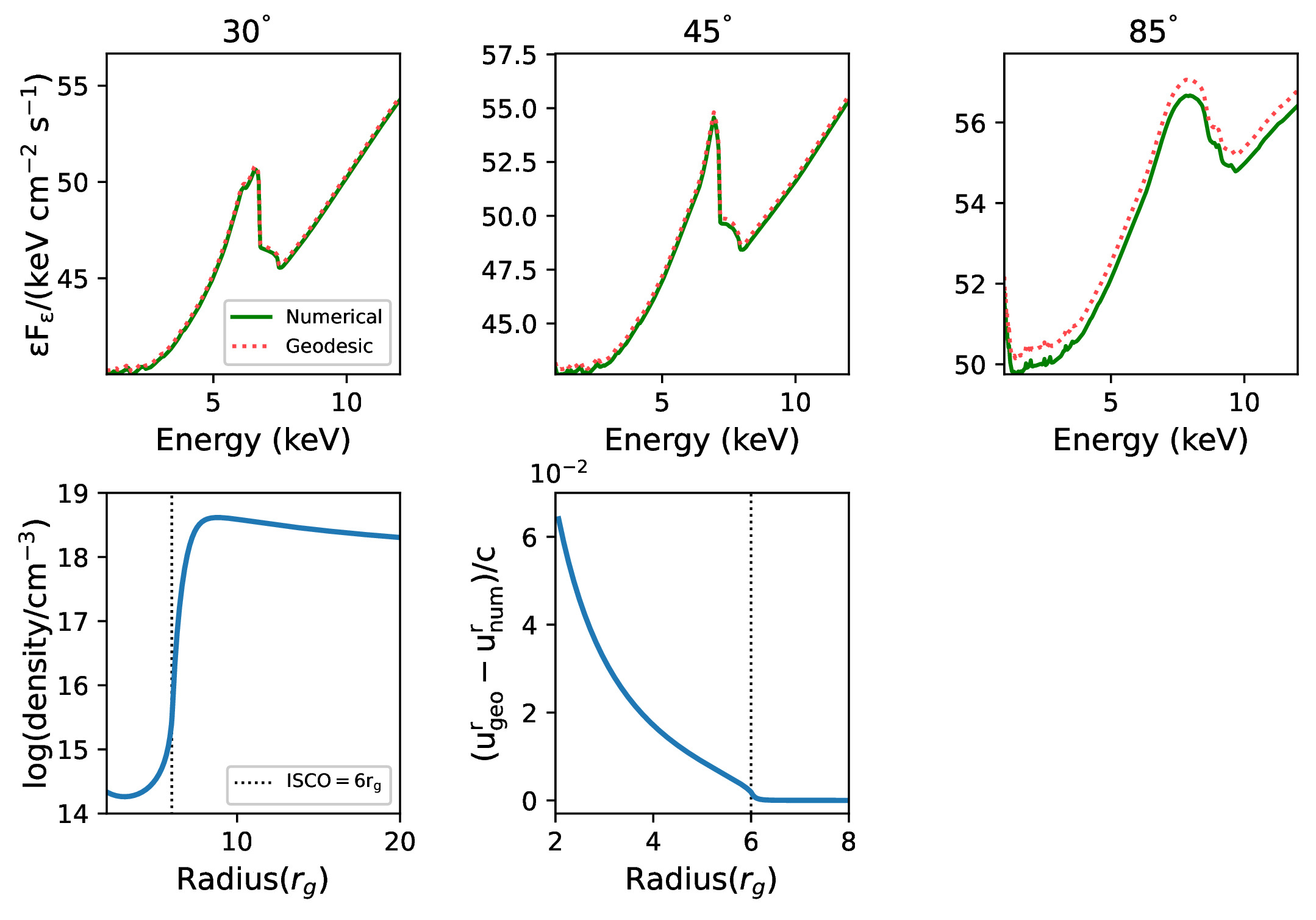}
\caption{Comparison of the $\rm Fe~K\alpha$ line spectral flux density from the numerical velocity profile and the time-like geodesic velocity profile. The second row shows the radial density profile and difference of two radial velocity profiles. The central black hole is Schwarzschild. The density profile and numerical velocity profile are from~\citep{Mummery2023}.} 
\label{fig:velocity}
\end{figure*}

As an initial comparison, for a Schwarzchild ($a=0$) black hole with ISCO stress $\delta_J=0.1$ we show the flux of the reflection spectrum close to the $\rm Fe~K\alpha$ line from the two cases (our geodesic profile as the red dashed line, with the numerical profile from \citep{Mummery2023} as a solid blue line) for a disk seen at three different inclinations in the upper panels of Fig.~\ref{fig:velocity}.  The lower two panels show the density profile from the numerical calculation, along with the difference in radial velocities between the two case. 

From this comparison we find a remarkable similarity;
%, as depicted in Fig.~\ref{fig:velocity}. 
the divergence is noticeable only at high inclination angles when the additional in-falling velocity has a more significant component along the observer's line-of-sight, marginally reducing the flux. This observation underscores the relative insignificance of $u^r_I$ in reflection spectroscopy, affirming the robustness of future spectral analysis against variations in geodesic and non-geodesic velocities. We can draw a preliminary observation that, in the Schwarzchild case, the discrepancy between geodesics and velocity profiles from~\citep{Mummery2023} is not important for the purpose of spectroscopy. 

\begin{figure*}
\includegraphics[width=\textwidth]{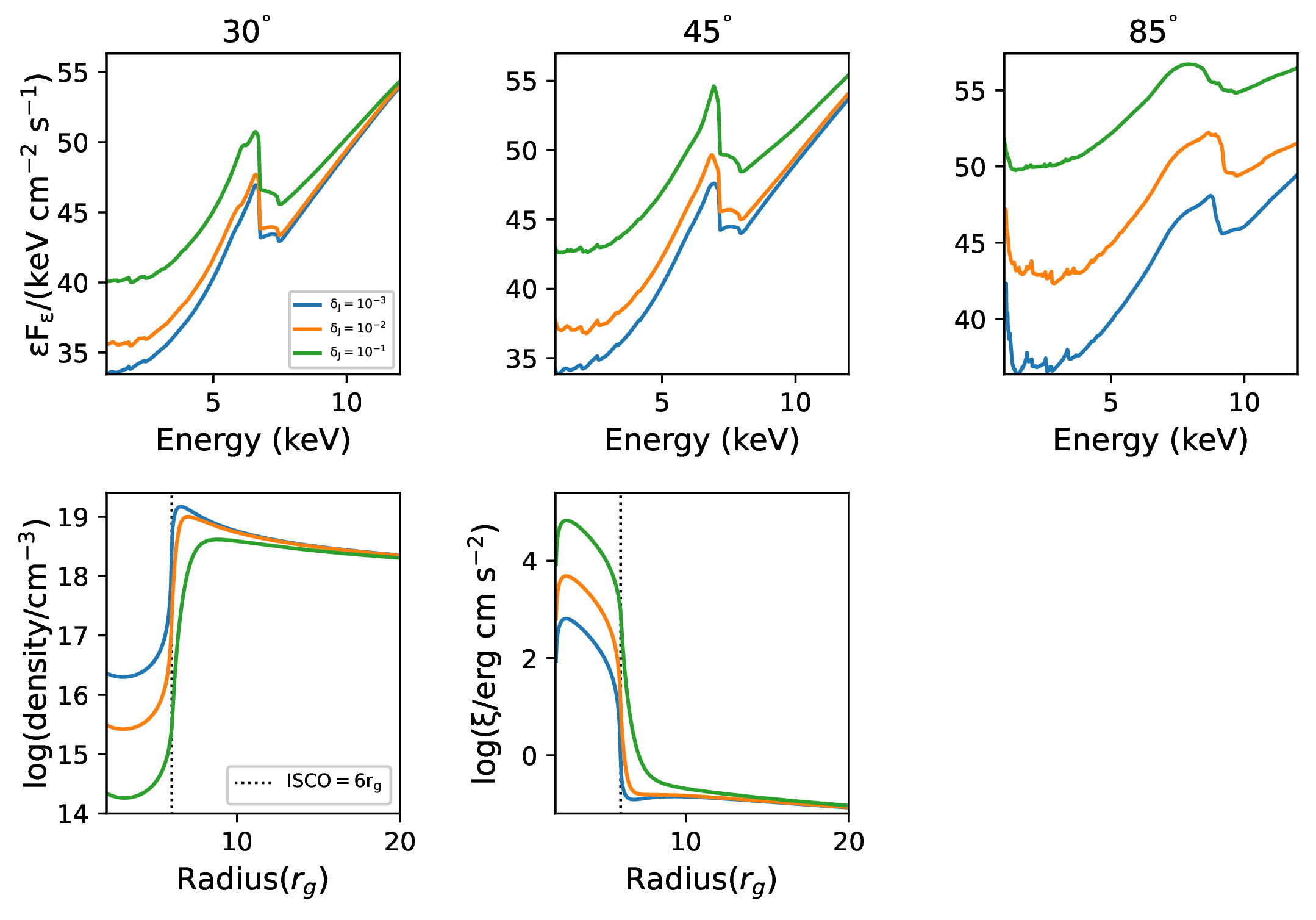}
\caption{The first row is a comparison of $\rm Fe~K\alpha$ line spectral flux density across various ISCO stress and inclination angle. The black hole is Schwarzschild. The second row depicts the density profile and ionization parameter resulted from different ISCO stress. For $\delta_J=0.1$, the ionization parameter at ISCO is set to be $\rm 10^{3} \, erg~cm~s^{-2}$. We change the ISCO ionization parameters of the $\delta_J=0.01, 0.001$ cases to keep $\epsilon(r_I)$ constant across all three cases.} 
\label{fig:Andy}
\end{figure*}

\subsubsection{Differentiating ISCO Stress of Generalized Novikov-Thorne Disks}
%As we mentioned in Section I, the ISCO stress, denoted as $\delta_J$, remains a free parameter in most thin disk models.
In most thin disk models, the ISCO stress, denoted as $\delta_J$, remains a free parameter.  
This condition assumes that there is no interaction between matter in the disk within the ISCO and outside the ISCO.
However, this condition is likely nonphysical (provided the ISCO is larger than the event horizon). Non-zero stress allows the transfer of angular momentum across the ISCO, affecting the velocity of the matter within the plunging region. In turn, the value of the ISCO stress affects density and velocity profiles of the entire disk (both within and outside the ISCO), and therefore affects the reflection spectrum we would see.

To see how the ISCO stress value changes the reflection spectrum,
we vary $\delta_J$ in the extended NT disk model of~\citep{Mummery2023}. From this we can make preliminary assessments on the potential of estimating the value of $\delta_J$ by modeling reflection spectra.

As examples, we consider disks with three values of the ISCO stress $\delta_J=10^{-3},10^{-2},10^{-1}$, as seen from three inclinations $30\degree,45\degree,85\degree$.  In the upper panels Fig.~\ref{fig:Andy}, we show the total reflected flux around the $\rm Fe~K\alpha$ line for these cases, while the lower panels show the disk density profile, and the ionization profile given the same illuminating flux $\epsilon(r)$. 

%By inspecting Fig.~\ref{fig:Andy}, we can make a few preliminary observations. 
From Fig.~\ref{fig:Andy}, we see that a higher ISCO stress causes a decrease in disk density in the inner regions of the disk and in the plunging region. Since the density decreases as $\delta_J$ increases, with the same $\epsilon(r)$, the ionization parameter also increases. It is noteworthy that, when $\delta_J=0.1$, the matter is almost completely ionized (for this particular choice of model parameters).

%hen the ISCO stress increases and density decreases, 
%the slope of the continuum between 1 -- 12 keV decreases. 
We see that the higher stress then causes the overall slope of reflected flux between 1 -- 12 keV to decrease. 
We also see that, across all inclinations, a higher ISCO stress further broaden the Fe $\rm K\alpha$ line profile. When seen from a low inclination, higher ISCO stress leads to a ``double-horned'' feature as introduced in~\citep{Muller2004}. At medium inclination, higher ISCO stress increases the contrast between the $\rm K\alpha$ line profile and its blue-side continuum. When observed at high inclination, a lower ISCO stress leads to stronger blue-shift of the $\rm Fe~K\alpha$ line.  

Overall, the increase in $\delta_J$ causes both an increase in the reflected flux in the 1 -- 12 keV energy band, along with shape changes to the $\rm Fe~K\alpha$ line which are large enough that they should be detectable with upcoming X-ray instruments such as XRISM~\citep{xrism}.

One important aspect of these calculations is that the density profile in Fig.~\ref{fig:Andy} approaches $10^{18}$$\rm cm^{-3}$ around 20$r_g$, which is 2 to 3 degrees of magnitudes larger than density suggested by standard accretion disk theory for a black hole mass of $10^7M_\odot$ \citep[e.g., see Fig.~1 in][]{density}. This is because the disk model in~\citep{Mummery2023} asymptotically matches Zone B (instead of Zone A) of the Shakura-Sunyaev disk~\citep{Shakura1973}, since Zone A is formally viscously unstable~\citep{instability}. At these high densities there are plasma effects that can change the atomic properties and thus affect the spectral profiles. Some of these effects have been included in the \texttt{XILLVER} reflection spectra used here \citep{density}, while additional effects have been recently incorporated by \citet{Ding2023}. We will incorporate these new \texttt{XILLVER} models in follow-up studies.
\label{deltaJ}
%\subsection{$\rm Fe~K\alpha$ Line Profile of Constant Density Disks}
%Since line profiles can be considered as $\delta$ functions of energy in their local frame, the line broadening effects seen by observer can be analyzed by the distribution of $g_{do}$ on the disk. 

%From Fig.~\ref{fig:K_Alpha}, we can observe that at 30$\degree$ inclination, the inclusion of the plunging region will only reduce the slope of the red wing of the $\rm Fe~K\alpha$ line, while the slope of the blue wing remains unchanged. We explain this by observing Fig.~\ref{fig:chop2}. The plunging region has the $g_{do}$ values generally smaller than the outer disk. And, at low inclinations, the plunging region has sufficient surface area with moderately small $g_{do}$ to make red wing contribution to the total flux. (Note that the $g_{do}^3$ effect discussed in Section~\ref{subsec:ChopDisk} will quickly suppress red wing contribution as $g_{do}$ decrease.) So the plunging region contributes a wider red wing to the total flux. Since at 30$\degree$, Doppler effect from the fluid velocity cannot outweigh gravitational red-shift, the plunging region has no contribution to the blue wing.

%\label{discuss::constant line profile}
\section{Conclusions}
In conclusion, our study advances the understanding of X-ray reflection in black holes, particularly focusing on the plunging regions of accretion disks. 

We first delve into the theoretical foundation for analyzing X-ray reflection in black holes. Section II focuses on the structure of radiation-disk interaction in the Kerr metric, discussing the system setup including the black hole, its accretion disk, and the corona. It covers topics such as the ISCO radius, velocity profiles and Lorentz factors of the accretion disk, and the interaction of light rays with the disk. This section lays the groundwork for understanding the complex dynamics at play in the vicinity of black holes, particularly in relation to X-ray reflection phenomena.

We also demonstrate that when the density profile remains constant or is sufficiently high within the ISCO to facilitate line formation, the reflected spectrum is significantly altered. Our findings corroborate previous research, highlighting a stronger impact of radiation reprocessed in the plunging region for black holes with lower spin. Additionally, the study confirms that relativistic effects, such as broadening of the iron line and beaming, vary with inclination, being more pronounced at low inclinations and leading to blueshift at high inclinations. This work contributes to a more nuanced understanding of accretion dynamics and spectral features around black holes.

We can also draw the following conclusions regarding the reflection from a generalized NT disk: (1) the reflection spectrum is robust with regards to the divergence between geodesic and non-geodesic velocity profiles from \citep{Mummery2023}; (2) with the various features identified in Section~\ref{deltaJ} and upcoming X-ray missions, we can differentiate the ISCO stress of black hole accretion disks.
%\section*{Acknowledgements}
% \begin{acknowledgments}
%This work was supported by a Leverhulme Trust International Professorship grant [number LIP-202-014].
% \end{acknowledgments}

%\appendix

%\section{Appendixes}

%\section{A little more on appendixes}

%\subsection{\label{app:subsec}A subsection in an appendix}

% The \nocite command causes all entries in a bibliography to be printed out
% whether or not they are actually referenced in the text. This is appropriate
% for the sample file to show the different styles of references, but authors
% most likely will not want to use it.
\nocite{*}

\bibliography{apssamp}% Produces the bibliography via BibTeX.

\end{document}